\documentclass[titlepage,11pt,final]{article}

\usepackage{amstext,amssymb,amsfonts,latexsym} 
\usepackage{theorem}
\usepackage{pifont}

 \renewcommand{\baselinestretch}{1.27}

 \def\bbox{\vrule height6pt width6pt depth1pt}

\theoremstyle{plain}
\theoremheaderfont{\bfseries}
 \newtheorem{theorem}{Theorem}[section]
 \newtheorem{lemma}[theorem]{Lemma}
 
 \newtheorem{corollary}[theorem]{Corollary}

 {\theorembodyfont{\rmfamily} \newtheorem{definition}[theorem]{Definition}}
 {\theorembodyfont{\rmfamily} } 
 {\theorembodyfont{\rmfamily} }

 \newenvironment{proof}{\par \noindent 
            {\bf Proof. \hs{2}}}{\hfill$\Box$ \vspace*{5mm}}

 \newcommand{\bs}{\bigskip}
 \newcommand{\ms}{\medskip}
 \newcommand{\n}{\noindent}
 \newcommand{\s}{\smallskip}
 \newcommand{\hs}[1]{\hspace*{ #1 mm}} 
 \newcommand{\vs}[1]{\vspace*{ #1 mm}}

 \newcommand{\integer}{\mathbb{Z}}
 \newcommand{\rational}{\mathbb{Q}}
 \newcommand{\complex}{\mathbb{C}}

 \newcommand{\algebraic}{\mathbb{A}}
 \newcommand{\ralgebraic}{\mathbb{A}\cap\mathbb{R}}

 \newcommand{\bm}[1]{\mbox{\boldmath $ #1 $}}

 \newcommand{\co}{\mathrm{co}\mbox{-}}

 \newcommand{\ie}{\textrm{i.e.},\hspace*{2mm}}

 \newcommand{\etalc}{\textrm{et al.}}

 \newcommand{\p}{\mathrm{{\bf P}}}
 \newcommand{\np}{\mathrm{{\bf NP}}}
 \newcommand{\bpp}{\mathrm{{\bf BPP}}}
 \newcommand{\pp}{\mathrm{{\bf PP}}}
 \newcommand{\ph}{\mathrm{{\bf PH}}}
 
 \newcommand{\cequalp}{\mathrm{{\bf C}}_{=}\mathrm{{\bf P}}}
 \newcommand{\app}{\mathrm{{\bf APP}}}

 \newcommand{\nqp}{\mathrm{{\bf NQP}}}
 \newcommand{\bqp}{\mathrm{{\bf BQP}}}

 \newcommand{\gapp}{\mathrm{{\bf GapP}}} 

 \newcommand{\pair}[1]{\langle #1 \rangle}
 \newcommand{\qbit}[1]{| #1 \rangle}
 \newcommand{\bra}[1]{\langle #1 |}
 \newcommand{\ket}[1]{| #1 \rangle}

\begin{document} 


\begin{center}
{\large {\bf NQP${}_{\complex}$ $=$ co\mbox{-}C$_{=}$P}} 
\ms

{Tomoyuki Yamakami\footnote{This work was supported by NSERC
Postdoctoral Fellowship and DIMACS Fellowship.} and Andrew
C. Yao\footnote{This work was supported in part by National Science
Foundation and DARPA under grant CCR-9627819.}} \ms

{\small 
{\em Department of Computer Science, Princeton University} \\
{\em Princeton, NJ 08544} 
}
\end{center}
\s

\centerline{{\bf Abstract}} \ms

{\small

Adleman, DeMarrais, and Huang introduced the nondeterministic quantum
polynomial-time complexity class $\nqp$ as an analogue of $\np$.
Fortnow and Rogers implicitly showed that, when the amplitudes are
rational numbers, $\nqp$ is contained in the complement of
$\cequalp$. Fenner, Green, Homer, and Pruim improved this result by
showing that, when the amplitudes are arbitrary algebraic numbers,
$\nqp$ coincides with $\co\cequalp$. In this paper we prove that, even
when the amplitudes are arbitrary complex numbers, $\nqp$ still
remains identical to $\co\cequalp$. As an immediate corollary, $\bqp$
differs from $\nqp$ when the amplitudes are unrestricted. \s

\n{\bf key words:} computational complexity, theory of computation

}

\section{Introduction}

In recent years, the possible use of the power of quantum interference
and entanglement to perform computations much faster than classical
computers has attracted attention from computer scientists and
physicists (e.g., \cite{Beni82,DJ92,Fey86,Gro96,Sch94,Simo97}).

In 1985 Deutsch \cite{Deu85} proposed the fundamental concept of {\em
quantum Turing machines} (see Bernstein and Vazirani \cite{BV97} for
further discussions). A quantum Turing machine is an extension of a
classical probabilistic Turing machine so that all computation paths
of the machine interfere with each other (similar to the phenomenon in
physics known as {\em quantum interference}). This makes it
potentially possible to carry out a large number of bit operations
simultaneously. Subsequent studies have founded the structural
analysis of quantum complexity classes.  In particular, quantum versus
classical counting computation has been a focal point in recent
studies {\cite{ADH97,FGHP98,FR98}}.

Adleman, DeMarrais, and Huang \cite{ADH97} introduced, as a quantum
analogue of $\np$, the ``nondeterministic'' quantum polynomial-time
complexity class $\nqp_{K}$, which is the set of decision problems
accepted with positive probability by polynomial-time quantum Turing
machines with amplitudes all drawn from set $K$. In their paper, they
argued that $\nqp_{\ralgebraic}$ lies within $\pp$, where $\algebraic$
denotes the set of algebraic complex numbers.

In classical complexity theory, Wagner \cite{Wag86} defined the
counting class $\cequalp$ as the set of decision problems that
determine whether the number of accepting computation paths (of
nondeterministic computation) equals that of rejecting computation
paths.  Fortnow and Rogers \cite{FR98} implicitly showed that
$\nqp_{\rational}\subseteq\co\cequalp$; in fact, as pointed out in
\cite{FGHP98}, their proof technique proves
$\nqp_{K}\subseteq\co\cequalp$ so long as all members of $K$ are
products of rational numbers and the square root of natural numbers.
Fenner, Green, Homer, and Pruim \cite{FGHP98} further improved this
result by showing $\nqp_{\algebraic}=\co\cequalp$, which gives a
characterization of $\nqp$ in terms of classical counting computation
when the amplitudes are restricted to algebraic numbers (in
\cite{FGHP98} $\nqp_{\algebraic}$ is succinctly denoted as
$\nqp$). Nevertheless, it has been unknown whether $\nqp_{\complex}$
further collapses to $\co\cequalp$.

In this paper we resolve this open question affirmatively as in
Theorem \ref{theorem:main}: $\nqp_{K}$ collapses to $\co\cequalp$ for
any set $K$ with $\rational\subseteq K\subseteq\complex$.  The proof
of the theorem consists of two steps. First we must show that
$\co\cequalp\subseteq\nqp_{\rational}$ (actually
$\co\cequalp\subseteq\nqp_{\{0,\pm\frac{3}{5},\pm\frac{4}{5},\pm1\}}$).
This claim was already mentioned in \cite{FGHP98} and its proof
recently appeared in \cite{FGHP98b}.  For completeness, we give its
proof in Section \ref{sec:main-rusult}. Then we must prove the claim
$\nqp_{\complex}\subseteq\co\cequalp$ in Section \ref{sec:main} by a
detailed algebraic analysis of transition amplitudes of quantum Turing
machines.

Our result yields another important consequence about the relationship
between $\nqp_{K}$ and $\bqp_{K}$, a quantum analogue of $\bpp$, which
was introduced by Bernstein and Vazirani \cite{BV97} as the set of
decision problems recognized by polynomial-time quantum Turing
machines with bounded-error with amplitudes from $K$. It is shown in
{\cite{ADH97}} that $\bqp_{\rational} =\bqp_{\algebraic}$ but
$\bqp_{\complex}$ has uncountable cardinality. Theorem
\ref{theorem:main} thus highlights a clear contrast between the power
of nondeterministic quantum computation and that of bounded-error
quantum computation: $\bqp_{\complex}\neq\nqp_{\complex}$. This
extends the separation of the exact quantum computation from
bounded-error quantum computation in the case when amplitudes are
unrestricted \cite{ADH97}.

The reader who needs more background on quantum computation may refer
to recent survey papers \cite{Ahar98,Bert97}.

\section{Basic Notions and Notation}

Let $\integer$ be the set of all integers, $\rational$ the set of
rational numbers, $\complex$ the set of complex numbers, and
$\algebraic$ the set of all algebraic complex numbers. Moreover, let
$\integer_{\geq0}$ and $\integer_{>0}$ denote the sets of all
nonnegative integers and of all positive integers, respectively. For
any $d\in\integer_{>0}$ and $k\in\integer_{\geq0}$, define
$\integer_{d}=\{i\in\integer\mid 0\leq i\leq d-1\}$ and
$\integer_{[k]}=\{i\in\integer\mid -k\leq i\leq k\}$.  By {\em
polynomials} we mean elements in $\integer[x_1,x_2,\ldots,x_m]$ for
some $m\in\integer_{\geq0}$.  For any finite sequence
$\bm{k}\in\integer^m$, let $|\bm{k}|=\max_{1\leq i\leq m}\{|k_i|\}$,
where $\bm{k}=(k_1,k_2,\ldots,k_m)$. Furthermore, $\bm{0}^k$ denotes
the $k$-tuple $(0,0,\ldots,0)$ for $k\in\integer_{>0}$.

Let $k\in\integer_{>0}$. A finite subset $\{\gamma_i\}_{1\leq i\leq
k}$ of $\complex$ is {\em linearly independent} if
$\sum_{i=1}^{k}a_i\gamma_i \neq0$ for any $k$-tuple
$(a_1,a_2,\ldots,a_k)\in \rational^k\setminus
\{\bm{0}^k\}$. Similarly, $\{\gamma_i\}_{1\leq i\leq k}$ is {\em
algebraically independent} if there is no $q$ in
$\rational[x_1,x_2,\ldots,x_k]$ such that $q$ is not identical to $0$
but $q(\gamma_1,\gamma_2,\ldots,\gamma_k)=0$.

We assume the reader's familiarity with classical complexity theory
and here we give only a brief description of quantum Turing machines
{\cite{BV97}}.  A $k$-track {\em quantum Turing machine} (QTM) $M$ is
a triplet $(\Sigma^k,Q,\delta)$, where $\Sigma$ is a finite alphabet
with a distinguished blank symbol $\#$, $Q$ is a finite set of states
with initial state $q_0$ and final state $q_f$, and $\delta$ is a
multi-valued {\em quantum transition function} from $Q\times\Sigma^k$
to $\complex^{Q\times\Sigma^k\times\{L,R\}}$.  A QTM has $k$ two-way
infinite tracks of cells indexed by $\integer$ and a read/write head
that moves along all the tracks. The expression
$\delta(p,\bm{\sigma},q,\bm{\tau},d)$ denotes the (transition)
amplitude in $\delta(p,\bm{\sigma})$ of
$\qbit{q}\qbit{\bm{\tau}}\qbit{d}$, where $\bm{\sigma},\bm{\tau}
\in\Sigma^k$ and $d\in\{L,R\}$.

A {\em superposition} of $M$ is a finite complex linear combination of
configurations of $M$ with unit $L_2$-norm. The {\em time-evolution
operator} of $M$ is a map from each superposition of $M$ to the
superposition of $M$ that results from a single application of the
transition function $\delta$. These time-evolution operators are
naturally identified with matrices.

The {\em running time} of $M$ on input $x$ is defined to be the
minimum integer $T$ such that, at time $T$, all computation paths of
$M$ on input $x$ have reached final configurations, and at any time
less than $T$ there are no final configurations, where a {\em final
configuration} is a configuration with state $q_f$.  We say that $M$
on input $x$ {\em halts in time $T$} if the running time of $M$ on
input $x$ is $T$.  The {\em final superposition} of $M$ is the
superposition that $M$ reaches when it halts. A QTM $M$ is called a
{\em polynomial-time QTM} if there exists a polynomial $p$ such that,
on every input $x$, $M$ halts in time $p(|x|)$.

A QTM is called {\em well-formed} if its time-evolution operator
preserves the $L_2$-norm. A QTM is {\em stationary} if it halts on all
inputs in a final superposition where each configuration has the head
in the start cells and a QTM is in {\em normal form} if, for every
$k$-tuple of track symbols $\bm{\sigma}$, $\delta(q_f,\bm{\sigma})=
\qbit{q_0}\qbit{\bm{\sigma}}\qbit{R}$.  For brevity, we say that a QTM
is {\em conservative} if it is well-formed and stationary and in
normal form. For any subset $K$ of $\complex$, we say that a QTM has
{\em $K$-amplitudes} if its transition amplitudes are all drawn from
$K$.

Let $M$ be a multitrack, well-formed QTM whose last track, called the
{\em output} track, has alphabet $\{0,1,\#\}$.  We say that $M$ {\em
accepts $x$ with probability} $p$ and also {\em rejects $x$ with
probability} $1-p$ if $M$ halts and $p$ is the sum of the squared
magnitudes of the amplitude of each final configuration in which the
output track consists only of $1$ as nonblank symbol in the start
cell. For convenience, we call such a final configuration an {\em
accepting configuration}.

For a more general model of quantum Turing machines, the reader may
refer to \cite{Yam99}. 
\section{Main Result}\label{sec:main-rusult}

In this section we state the main theorem of this paper. First 
we give the formal definitions of the complexity classes $\cequalp$
\cite{Wag86} and $\nqp$ \cite{ADH97}.

The counting class $\cequalp$ was originally introduced by Wagner
\cite{Wag86}. For convenience, we begin with the definition of
$\gapp$-functions. For a nondeterministic Turing machine $M$,
$Acc_M(x)$ denotes the number of accepting computation paths of $M$ on
input $x$. Similarly, we denote by $Rej_M(x)$ the number of rejecting
computation paths of $M$ on input $x$.

\begin{definition}{\rm {\bf \cite{FFK94}}}\hs{1}
A function from $\Sigma^*$ to $\integer$ is in $\gapp$ if there exists
a polynomial-time nondeterministic Turing machine $M$ such that
$f(x)=Acc_{M}(x)-Rej_{M}(x)$ for every string $x$.
\end{definition}

\begin{lemma}\label{lemma:property}{\rm {\bf \cite{FFK94}}}\hs{1}
Let $f\in\gapp$ and $p$ a polynomial. Then, the following functions
are also $\gapp$-functions: $f^2$, $\lambda
x.\sum_{y\in\Sigma^{p(|x|)}}f(x,y)$, and $\lambda
x.\prod_{i=1}^{p(|x|)}f(x,1^i)$, where $f^2(x)$ means $(f(x))^2$ and
the $\lambda$-notation $\lambda x.g(x)$ means the function $g$.
\end{lemma}

\begin{definition}\label{def:cequalp}{\rm {\bf \cite{Wag86}}}\hs{1}
A set $S$ is in $\cequalp$ if there exists a $\gapp$-function $f$ such
that, for every $x$, $x\in S$ exactly when $f(x)=0$.
\end{definition}

Adleman, DeMarrais, and Huang \cite{ADH97} introduced the notion of
``nondeterministic'' quantum computation and defined the complexity
class $\nqp_{K}$ as the collection of all sets that can be recognized
by nondeterministic QTMs with $K$-amplitudes in polynomial time.

\begin{definition}\label{def:nqp}{\rm {\bf \cite{ADH97}}}\hs{1}
Let $K$ be a subset of $\complex$. A set $S$ is in $\nqp_{K}$ if there
exists a polynomial-time, conservative QTM $M$ with $K$-amplitudes
such that, for every $x$, if $x\in S$ then $M$ accepts $x$ with
positive probability and if $x\not\in S$ then $M$ rejects $x$ with
probability 1.
\end{definition}

It immediately follows from Definition \ref{def:nqp} that
$\np\subseteq\nqp_{\rational}\subseteq\nqp_{\algebraic}
\subseteq\nqp_{\complex}$.  Adleman \etalc \cite{ADH97} first showed
that $\nqp_{\ralgebraic}$ is a subset of $\pp$. Based on the work of
Fortnow and Rogers \cite{FR98}, Fenner, Green, Homer, and Pruim
\cite{FGHP98} later obtained the significant improvement:
$\nqp_{K}=\co\cequalp$ for any set $K$ satisfying $\rational\subseteq
K\subseteq\algebraic$.

We expand their result and show as the main theorem that any class
$\nqp_{K}$, $\rational \subseteq K\subseteq\complex$, collapses to
$\co\cequalp$.

\begin{theorem}\label{theorem:main}
For any set $K$ with $\rational\subseteq K\subseteq\complex$,
$\nqp_{K}=\co\cequalp$.
\end{theorem}

Before giving the proof of Theorem \ref{theorem:main}, we state its
immediate corollary. We need the notion of bounded-error quantum
polynomial-time complexity class $\bqp_{K}$ given by Bernstein and
Vazirani \cite{BV97}.

\begin{definition}\label{def:bqp}{\rm {\bf \cite{BV97}}}\hs{1}
A set $S$ is in $\bqp_{K}$ if there exists a polynomial-time,
conservative QTM $M$ with $K$-amplitudes such that, for every $x$, if
$x\in S$ then $M$ accepts $x$ with probability at least $\frac{2}{3}$
and if $x\not\in S$ then $M$ rejects $x$ with probability at least
$\frac{2}{3}$.
\end{definition}

It is known from \cite{ADH97} that $\bqp_{\complex}$ has uncountable
cardinality. Theorem \ref{theorem:main} thus implies that
$\bqp_{\complex}$ differs from $\nqp_{\complex}$.

\begin{corollary}
$\bqp_{\complex}\neq\nqp_{\complex}$.
\end{corollary}

The proof of Theorem \ref{theorem:main} consists of two parts:
$\co\cequalp\subseteq\nqp_{\{0,\pm\frac{3}{5},\pm\frac{4}{5},\pm1\}}$
and $\nqp_{\complex}\subseteq\co\cequalp$. The proof of the first
claim recently appeared in \cite{FGHP98b}. For completeness, however,
we include a proof of the first claim below. The second claim needs an
elaborate argument and will be proved in the next section.

Let $S$ be any set in $\co\cequalp$. By definition, there exists a
$\gapp$-function $f$ such that, for every $x$, $x\in S$ if and only if
$f(x)\neq0$. Without loss of generality, we can assume that, for some
polynomial $p$ and some deterministic polynomial-time computable
predicate\footnote{ A predicate can be seen as a function from
$\{0,1\}^*\times \{0,1\}^*$ to $\{0,1\}$.} $R$,
$f(x)=|\{y\in\{0,1\}^{p(|x|)}\mid R(x,y)=1\}|-
|\{y\in\{0,1\}^{p(|x|)}\mid R(x,y)=0\}|$ for all binary strings $x$.

We wish to design a quantum algorithm with
$\{0,\pm\frac{3}{5},\pm\frac{4}{5},\pm1\}$-amplitude that produces, on
input $x$, a particular configuration with amplitude
$-\epsilon^{p(|x|)+1}f(x)$, where $\epsilon=12/25$, so that we can
observe this configuration with positive probability if and only if
$x\in S$. This implies that $S$ is in
$\nqp_{\{0,\pm\frac{3}{5},\pm\frac{4}{5},\pm1\}}$.  To simplify our
argument, we make use of the four letter alphabet
$\Sigma_4=\{0,1,2,3\}$.

Let $I$ be the identity transform and let $H[a,b|\delta]$ be the
generalized Hadamard transform defined as
$\sum_{y,u\in\{a,b\}}(-1)^{[y=u=b]}\delta^{[y=u]}(1-\delta)^{[y\neq
u]}\ket{u}\bra{y}$, where $a,b\in\Sigma_4$, $\delta\in\complex$, and
the square brackets mean the truth value.\footnote{Conventionally we
set TRUTH=1 and FALSE=0. For example, $[0=0]=1$ and $[0=1]=0$.}
Moreover, let $H= H[0,1|\frac{4}{5}]$, $J= H[0,1|\frac{3}{5}]$, and
$K=H[0,2|\frac{3}{5}]+ H[1,3|\frac{4}{5}]$.  Notice that $H$, $I$,
$J$, and $K$ are unitary and their amplitudes are all in
$\{0,\pm\frac{3}{5},\pm\frac{4}{5},\pm1\}$.

Let $x$ be an input of length $n$.  We start with the initial
superposition $\qbit{\phi_0}=\qbit{0^{p(n)}}\qbit{0}$.  We apply the
operations $H^{p(n)}\otimes I$ to $\qbit{\phi_0}$. Next we change the
content of the last track from $\qbit{0}$ to $\qbit{R(x,y)}$. This can
be done reversibly in polynomial time since $R$ is computable by a
polynomial-time reversible Turing machine \cite{Ben73,BV97}. Finally
we apply the operations $J^{p(n)}\otimes JK$ to this last
superposition and let $\qbit{\phi}$ denote the consequence.


Let $\qbit{\phi_1}$ denote the {\em observable}
$\qbit{0^{p(n)}}\qbit{1}$. When we observe $\qbit{\phi}$, we can find
state $\qbit{\phi_1}$ with amplitude $\langle{\phi_1}|{\phi}\rangle$,
which is
$\epsilon^{p(n)}\sum_{y\in\{0,1\}^{p(n)}}(-1)^{R(x,y)}\epsilon$ since
$\bra{1}JK\ket{R(x,y)}=(-1)^{R(x,y)}\epsilon$. By the definition of
$f$, $\langle{\phi_1}|{\phi}\rangle$ is equal to
$-\epsilon^{p(n)+1}f(x)$.

\section{Proof of the Main Theorem}\label{sec:main}

This section completes the proof of Theorem \ref{theorem:main} by
proving $\nqp_{\complex}\subseteq\co\cequalp$.  Assume that $S$ is in
$\nqp_{\complex}$.  By Definition \ref{def:nqp}, there exists an
element $p\in\integer[x]$ and an $\ell$-track conservative quantum
Turing machine $M=(\Sigma,Q,\delta)$ with $\complex$-amplitudes that
recognizes $S$ in time $p(n)$ on any input of length $n$. Let $D$ be
the set of all transition amplitudes of $\delta$; that is,
$D=\{\delta(p',\bm{\sigma},q',\bm{\tau},d')\mid p',q'\in Q,
\bm{\sigma},\bm{\tau}\in\Sigma^{\ell}, d'\in\{L,R\}\}$.  We must show
that $S$ is in $\co\cequalp$.

The key ingredient of our proof is, similar to Lemma 6.6 in
\cite{ADH97}, to show that, for some constant $u\in\complex$, every
amplitude of a configuration in a superposition generated by $M$ at
time $t$, when multiplied by the factor $u^{2t-1}$, is uniquely
expressed as a linear combination of $O(poly(t))$ linearly independent
monomials with integer coefficients. If each basic monomial is
properly indexed, any transition amplitude can be encoded as a
collection of pairs of such indices and their integer
coefficients. This encoding enables us to carry out amplitude
calculations on a classical Turing machine.

We first show that any number in $D$ can be expressed in a certain
canonical way.  Let $A=\{\alpha_i\}_{1\leq i\leq m}$ be any maximal
algebraically independent subset of $D$ and define $F=\rational(A)$,
{\ie} the field generated by all elements in $A$ over $\rational$.  We
further define $G$ to be the field generated by all the elements in
$\{1\}\cup(D\setminus A)$ over $F$.  Let $B=\{\beta_i\}_{0\leq i< d}$
be a basis of $G$ over $F$. For convenience, we assume $\beta_0=1$ so
that, even in the special case $A=D$, $\{\beta_0\}$ becomes a basis of
$G$ over $F$. Let $D'=D\cup \{\beta_i\beta_j\}_{0\leq i,j< d}$.

For each element $\alpha$ in $G$, since $B$ is a basis, $\alpha$ can
be uniquely written as $\sum_{j=0}^{d-1}\lambda_{j}\beta_j$ for some
$\lambda_{j}\in F$. Since the elements in $A$ are all algebraically
independent, each $\lambda_j$ can be written as $s_j/u_j$, where each
$s_j$ and $u_j$ is a finite sum of linearly independent monomials of
the form $a_{\bm{{}_k}_j} (\prod_{i=1}^m\alpha_i^{k_{ij}})$ for some
$\bm{k}_j=(k_{1j},k_{2j},\ldots,k_{mj})\in\integer^m$ and
$a_{\bm{{}_k}_j}\in\integer$. Unfortunately, this representation is in
general not unique, since $s_j/u_j = (s_jr)/(u_jr)$ for any non-zero
element $r$.

To give a standard form for all the elements in $D'$, we need to
``normalize'' them by choosing an appropriate common denominator. Let
$u\in G$ be any common denominator of all the elements $\alpha$ in
$D'$ such that $u\alpha$ is written as
$\sum_{\bm{{}_k}}a_{\bm{{}_k}}(\prod_{i=1}^{m}\alpha_i^{k_i})\beta_k$,
where $\bm{k}=(k,k_1,k_2,\ldots,k_m)\in\integer_d\times \integer^m$
and $a_{\bm{{}_k}}\in\integer$. Notice that such a form is uniquely
determined by a collection of pairs of $\bm{k}$ and $a_{\bm{{}_k}}$.
We call this unique form the {\em canonical form} of $u\alpha$. Fix
$u$ from now on.  For the canonical form, we call $\bm{k}$ an {\em
index} and $a_{\bm{{}_k}}$ a {\em major sign} of $u\alpha$ with
respect to index $\bm{k}$ (or a {\em major $\bm{k}$-sign}, for
short). An index $\bm{k}$ is said to be {\em principal} if the major
$\bm{k}$-sign is nonzero. For each $\alpha\in D'$, let $ind(u\alpha)$
be the maximum of $|\bm{k}|$ over all principal indices $\bm{k}$ of
$u\alpha$.  Moreover, let $e$ be the maximum of $d$ and of
$ind(u\alpha)$ over all elements $\alpha$ in $D'$.

A crucial point of our proof relies on the following lemma.

\begin{lemma}\label{lemma:index}
The amplitude of each configuration of $M$ on input $x$ in any
superposition at time $t$, $t>0$, when multiplied by the factor 
$u^{2t-1}$, can be written in the canonical form
$\sum_{\bm{{}_k}}a_{\bm{{}_k}}(\prod_{i=1}^{m}\alpha_i^{k_i})\beta_k$,
where $\bm{k}=(k,k_1,k_2,\ldots,k_m)$ ranges over $\integer_d
\times(\integer_{[2et]})^m$ and $a_{\bm{{}_k}}\in\integer$.
\end{lemma}

\begin{proof}
Let $\alpha_{C,t}$ denote the amplitude of configuration $C$ of $M$ on
input $x$ in a superposition at time $t$. When $t=1$, the lemma is
trivial. Let $C'$ be any configuration in a superposition at time
$t+1$. Note that $u^{2t+1}\alpha_{C',t+1}$ is a sum of
$u^2(u^{2t-1}\alpha_{C,t})\delta_{C,C'}$ over all configurations $C$,
where $\delta_{C,C'}$ denotes the transition amplitude of $\delta$
that corresponds to the transition from $C$ to $C'$ in a single step.
By the induction hypothesis, $u^{2t-1}\alpha_{C,t}$ has a canonical
form as in the lemma. Hence, it suffices to show that, for each
configuration $C$ and each index $\bm{k}\in
\integer_d\times(\integer_{[2et]})^m$, $\alpha'_{C,C',\bm{{}_k}}
\stackrel{\mathrm{def}}{=} u^2(\prod_{i=1}^m\alpha_i^{k_i})
\beta_k\delta_{C,C'}$ has a canonical form in which all the principle
indices lie in $\integer_d\times(\integer_{[2e(t+1)]})^m$ since
$u^{2t+1}\alpha_{C',t+1}$ is expressed as the sum of
$a_{\bm{{}_k}}\alpha'_{C,C',\bm{{}_k}}$ over all $C$ and $\bm{k}$.

Let $\bm{k}=(k,k_1,\ldots,k_m)$ be an index in
$\integer_d\times(\integer_{[2et]})^m$, which corresponds to monomial
$(\prod_{i=1}^m\alpha_i^{k_i})\beta_k$.  We first show that
$\alpha'_{C,C',\bm{{}_k}}$ has a canonical form. Since $\delta_{CC'}\in
D'$, we can assume that the canonical form of $u\delta_{C,C'}$ is
$\sum_{\bm{{}_j}} b_{\bm{{}_j}}
(\prod_{i=1}^m\alpha_i^{j_i})\beta_{j}$, where
$\bm{j}=(j,j_1,\ldots,j_m)$ ranges over
$\integer_d\times(\integer_{[e]})^{m}$ and $b_{\bm{{}_j}}\in\integer$.
Then, $\alpha'_{C,C',\bm{{}_k}}$ is written as: \bs

$(*)$ \hs{10}
$\alpha'_{C,C',\bm{{}_k}} 
= \sum_{\bm{{}_j}}b_{\bm{{}_j}} 
 \left(\prod_{i=1}^m\alpha_i^{k_i+j_i}\right)u\beta_{k}\beta_{j} 
= \sum_{\bm{{}_j}}\sum_{\bm{{}_h}_{j}}
 b_{\bm{{}_j}}c_{\bm{{}_h}_{j}}
 \left(\prod_{i=1}^m\alpha_i^{k_i+j_i+h_{ij}}\right)\beta_{h_{j}}$,
\bs

\n provided that $u\beta_{k}\beta_{j}$ has a canonical form
$\sum_{\bm{{}_h}_{j}}c_{\bm{{}_h}_{j}}
(\prod_{i=1}^m\alpha_i^{h_{ij}})\beta_{h_{j}}$, where
$\bm{h}_{j}=(h_{j},h_{1j},\ldots,h_{mj})$ ranges over
$\integer_d\times(\integer_{[e]})^{m}$ and
$c_{\bm{{}_h}_{j}}\in\integer$.  Since
$b_{\bm{{}_j}}c_{\bm{{}_h}_{j}}\in\integer$, $\alpha'_{C,C',\bm{{}_k}}$
must have a canonical form. For later use, let
$h(x,C,\bm{k},C',\bm{k'})$ be the major $\bm{k'}$-sign of
$\alpha'_{C,C',\bm{{}_k}}$ for any index $\bm{k'}=(k',k'_1,\ldots,k'_m)$.

We next show that $ind(\alpha'_{C,C',\bm{{}_k}})\leq 2e(t+1)$.  By $(*)$
it follows that $ind(\alpha'_{C,C',\bm{{}_k}})$ is bounded above by the
maximum of $k_i+j_i+h_{ij}$, which is at most
$|\bm{k}|+|\bm{j}|+|\bm{h}_j|\leq 2et + 2e = 2e(t+1)$; in other words,
all the principal indices of $\alpha'_{C,C',\bm{{}_k}}$ must lie in
$\integer_d\times(\integer_{[2e(t+1)]})^{m}$. This also shows that
$h(x,C,\bm{k},C',\bm{k'})$ is computed\footnote{We assume that $C$,
$C'$, $\bm{k}$, and $\bm{k'}$ are appropriately encoded into strings
in $\Sigma^*$.} deterministically in time polynomial in the length of
$C$ and $C'$ and also in $|\bm{k}|$ and $|\bm{k'}|$ since
$h(x,C,\bm{k},C',\bm{k'})$ is the sum of
$b_{\bm{{}_j}}c_{\bm{{}_h}_{j}}$ over all pairs of $\bm{j}$ and
$\bm{h}_j$ such that $h_j=k'$ and $k_i+j_i+h_{ij}=k'_i$ for each $i$
with $1\leq i\leq m$.
\end{proof}

In what follows, we show how to simulate a quantum computation of $M$.
First we define a function $f$ as follows. Let $x$ be a string of
length $n$, $C$ an accepting configuration of $M$ on input $x$, and
$\bm{k}$ an index. Let $f(x,C,\bm{k})$ be the major $\bm{k}$-sign of
$u^{2p(n)-1}$ times the amplitude of $C$ in the final superposition of
$M$ on input $x$.  For convenience, we set $f(x,C,\bm{k})=0$ for any
other set of inputs $(x,C,\bm{k})$. 

Notice by Lemma \ref{lemma:index} that $M$ rejects $x$ with certainty
if and only if the amplitude of any accepting configuration of $M$ on
$x$, multiplied by $u^{2p(n)-1}$, has major sign 0 with respect to any
principal index. The following lemma is thus immediate.

\begin{lemma}\label{lemma:sum}
For every $x$, $x\not\in S$ if and only if, for every accepting
configuration $C$ of $M$ on input $x$ and for every index
$\bm{k}\in\integer_d\times(\integer_{[2ep(n)]})^m$, $f(x,C,\bm{k})=0$.
\end{lemma}

We want to show that $f$ is a $\gapp$-function.  Theorem
\ref{theorem:main} follows once this is proved.  To see this, define
\[
 g(x)= \sum_{C}\sum_{\bm{{}_k}}f^2(x,C,\bm{k}),
\]
where $C$ ranges over all accepting configurations of $M$ on input $x$
and $\bm{k}$ is drawn from $\integer_d\times(\integer_{[2ep(n)]})^m$.
It follows from Lemma \ref{lemma:property} that $g$ is also in
$\gapp$, and by Lemma \ref{lemma:sum} $g(x)=0$ if and only if $x\not\in
S$. This yields the desired conclusion that $S$ is in $\co\cequalp$.

To show $f\in\gapp$, let $\bm{C}=\pair{C_0,C_1,\ldots,C_{p(n)}}$ be
any ``computation path'' of $M$ on input $x$ of length $n$; that is,
$C_0$ is the initial configuration of $M$ on input $x$ and $\delta$
transforms $C_{i-1}$ into $C_{i}$ in a single step. Also let
$\bm{K}=\pair{\bm{k}_0,\bm{k}_1,\ldots,\bm{k}_{p(n)}}$ be any sequence
of indices in $\integer_d\times(\integer_{[2ep(n)]})^m$ such that
$\bm{k}_0=\bm{0}^{m+1}$. We define $h'(x,\bm{C},\bm{K})$ to be the
product of $h(x,C_{i-1},\bm{k}_{i-1},C_i,\bm{k}_i)$ over all $i$,
$1\leq i\leq p(n)$, where $h$ has been defined in the proof of Lemma
\ref{lemma:property}.  Notice that $h'$ is polynomial-time computable
since $h$ is. 

Note that the sum $\sum_{\bm{{}_K}} h'(x,\bm{C},\bm{K})$ relates to
the major $\bm{k}$-sign of the amplitude, multiplied by $u^{2p(n)-1}$,
of the computation path $\bm{C}$, where
$\bm{K}=\pair{\bm{k}_0,\bm{k}_1,\ldots,\bm{k}_{p(n)}}$ ranges over
$\integer_d\times(\integer_{[2ep(n)]})^m$ with $\bm{k}_0=\bm{0}^{m+1}$
and $\bm{k}_{p(n)}=\bm{k}$.  The following equation is thus
straightforward.
\[
 f(x,C,\bm{k}) = \sum_{\bm{{}_K}}\sum_{\bm{{}_C}}h'(x,\bm{C},\bm{K}),
\]
where $\bm{K}=\pair{\bm{k}_0,\bm{k}_1,\ldots,\bm{k}_{p(n)}}$ ranges
over $(\integer_d\times(\integer_{[2ep(n)]})^m)^{p(n)}$ and
$\bm{C}=\pair{C_0,C_1,\ldots,C_{p(n)}}$ is a computation path of $M$
on input $x$ such that $\bm{k}_0=\bm{0}^{m+1}$,
$\bm{k}_{p(n)}=\bm{k}$, and $C_{p(n)}=C$. 

Lemma \ref{lemma:property} guarantees that $f$ is indeed a
$\gapp$-function.  This completes the proof of Theorem
\ref{theorem:main}.

\section{Discussion}

We have extended earlier works of \cite{ADH97,FGHP98,FR98} to show
that nondeterministic polynomial-time quantum computation with
arbitrary amplitudes can be completely characterized by Wagner's
polynomial-time counting computation. Our result thus makes it
possible to define the class $\nqp$ independent of the choice of
amplitudes, whereas $\bqp_{\complex}$ is known to differ from
$\bqp_{\rational}$ \cite{ADH97}.  We also note that the proof of
Theorem \ref{theorem:main} can relativize to an arbitrary oracle $A$;
namely, $\nqp_{K}^{A}=\co\cequalp^{A}$ for any set $K$ with
$\rational\subseteq K\subseteq\complex$. As a result, for instance, we
have $\nqp^{\nqp}=\co\cequalp^{\cequalp}$ and thus
$\nqp\subseteq\pp\subseteq\nqp^{\nqp}\subseteq\pp^{\pp}$. This implies
that the hierarchy built over $\nqp$, analogous to the polynomial-time
hierarchy, interweaves into Wagner's counting hierarchy \cite{Wag86}
over $\pp$.

At the end, we remind the reader that the fact $\nqp=\co\cequalp$
yields further consequences based on the well-known results on the
class $\cequalp$. For example, $\pp^{\ph}\subseteq\np^{\nqp}$ follows
directly from $\pp^{\ph}\subseteq\p^{\pp}$ \cite{Toda89} and
$\np^{\pp}=\np^{\cequalp}$ \cite{Tor91} and it also follows from
\cite{Li93} that all sparse $\nqp$ sets are in $\app$. Moreover,
$\nqp=\co\nqp$ if and only if $\ph^{\pp}=\nqp$, which follows from a
result in \cite{Gre93}. Note that these results also follow from
\cite{FGHP98}.

 \renewcommand{\baselinestretch}{1}

\bibliographystyle{alpha}

\begin{thebibliography}{Gur91}
{\small

\bibitem{ADH97}
L. M. Adleman, J. DeMarrais, and M. A. Huang, Quantum computability,
{\em SIAM J. Comput.}, {\bf 26} (1997), 1524--1540.
\vs{-2}
\bibitem{Ahar98}
D. Aharonov, Quantum computation, in {\em Annual Reviews of
Computational Physics VI}, ed. Dietrich Stauffer, World Scientific,
1998.
\vs{-2}
\bibitem{Ben73}
C. H. Bennett, Logical reversibility of computation, {\em IBM
J. Res. Develop.}, {\bf 17} (1973), 525--532.
\vs{-2}
\bibitem{BV97}
E. Bernstein and U. Vazirani, Quantum complexity theory, {\em SIAM
J. Comput.}, {\bf 26} (1997), 1411--1473.
\vs{-2}
\bibitem{Bert97}
A. Berthiaume, Quantum computation, in {\em Complexity Theory
Retrospective II}, eds. L.A. Hemaspaandra and A.L. Selman, pp.23--51, 
Springer, 1997.
\vs{-2}
\bibitem{Beni82}
P. Benioff, Quantum mechanical Hamiltonian models of Turing machines,
{\em J. Stat. Phys.} {\bf 29} (1982), 515--546.
\vs{-2}
\bibitem{Deu85} 
D. Deutsch, Quantum theory, the Church-Turing principle and the
universal quantum computer, {\em Proc. Roy. Soc. London,} Ser.A, {\bf
400} (1985), 97--117.  
\vs{-2}
\bibitem{Deu89} 
D. Deutsch, Quantum computational networks, {\em
Proc. Roy. Soc. London,} Ser.A, {\bf 425} (1989), 73--90.
\vs{-2}
\bibitem{DJ92} 
D. Deutsch and R. Jozsa, Rapid solution of problems by quantum 
computation, {\em Proc. Roy. Soc. London,} Ser.A, {\bf 439} (1992),
553--558.
\vs{-2}
\bibitem{FFK94}
S. Fenner, L. Fortnow, and S. Kurtz, Gap-definable counting classes,
{\em J. Comput. and System Sci.}, {\bf 48} (1994), 116--148.
\vs{-2}
\bibitem{FGHP98}
S. Fenner, F. Green, S. Homer, and R. Pruim, Quantum NP is hard for
PH, in {\em Proc. 6th Italian Conference on Theoretical Computer
Science}, World-Scientific, Singapore, pp.241--252, 1998.
\vs{-2}
\bibitem{FGHP98b}
S. Fenner, F. Green, S. Homer, and R. Pruim, Determining acceptance
possibility for a quantum computation is hard for the polynomial
hierarchy, quant-ph/9812056, December 18, 1998.
\vs{-2}
\bibitem{Fey86}
R. Feynman, Quantum mechanical computers, {\em Found.
Phys.}, {\bf 16} (1986), 507--531.
\vs{-2}
\bibitem{FR98}
L. Fortnow and J. Rogers, Complexity limitations on quantum
computation,
{\em Proc. 13th IEEE Conference on Computational Complexity},
pp.202--209, 1998.
\vs{-2}
\bibitem{Gre93}
F. Green, On the power of deterministic reductions to C${}_=$P, {\em
Math. Systems Theory}, {\bf 26} (1993), 215--233.
\vs{-2}
\bibitem{Gro96}
L. K. Grover, A fast quantum mechanical algorithm for database search,
{\em Proceedings of 28th ACM Symposium on Theory of Computing},
pp.212-219, 1996.  
\vs{-2}
\bibitem{Li93}
L. Li, {\em On the counting functions}, Ph.D. dissertation, Department of
Computer Science, University of Chicago, 1993.
\vs{-2}
\bibitem{Sch94}
P. W. Shor, Polynomial-time algorithms for prime factorization and
discrete logarithms on a quantum computer, {\em SIAM J. Comput.}, {\bf
26} (1997), 1484--1509.
\vs{-2}
\bibitem{Simo97}
D. R. Simon, On the power of quantum computation, {\em SIAM J. Comput},
{\bf 26} (1997), 1474--1483.  
\vs{-2}
\bibitem{Toda89}
S. Toda, PP is as hard as the polynomial-time hierarchy, {\em SIAM
J. Comput.}, {\bf 20} (1991), 865--877.
\vs{-2}
\bibitem{Tor91}
J. Tor{\'a}n, Complexity classes defined by counting quantifiers, {\em
J. ACM}, {\bf 38} (1991), 753--774.
\vs{-2}
\bibitem{Wag86}
K. Wagner, The complexity of combinatorial problems with succinct
input representation, {\em Acta Inf.} {\bf 23} (1986), 325--356.
\vs{-2}
\bibitem{Yam99}
T. Yamakami, A foundation of programming a multi-tape quantum Turing
machine, to appear in {\em Proc. 24th International Symposium on
Mathematical Foundation of Computer Science}, Lecture Note in Computer
Science, September, 1999.
}
\end{thebibliography}


\end{document}